\documentclass[manuscript]{aastex}

\slugcomment{Not to appear in Nonlearned J., 45.}

\shorttitle{Ni I lines in the solar spectrum}
\shortauthors{Vieytes et al.}

\begin{document}


\title{Improving the Ni I atomic model for solar and stellar atmospheric models 
    }


\author{M.C. Vieytes}
\affil{Instituto de de Astronom\'ia y F\'isica del Espacio, CONICET and UNTREF, Buenos Aires, Argentina}

\and

\author{J.M. Fontenla}
\affil{Laboratory for Atmospheric and Space Physics, University of Colorado, Boulder, USA}
\email{mariela@iafe.uba.ar;johnf@digidyna.com}



\begin{abstract}
Neutral nickel (Ni I) is abundant in the solar atmosphere and is one of the important elements that contribute to the emission and absorption of radiation in the spectral range between 1900 and 3900 \AA. Previously, the Solar Radiation Physical Modeling (SRPM) models of the solar atmosphere considered only few levels of this species.
Here we improve the Ni I atomic model by taking into account 61 levels and 490 spectral lines. We compute the populations of these levels in full NLTE using the SRPM code and compare the resulting emerging spectrum with observations.

The present atomic model improves significantly the calculation of the solar spectral irradiance at near-UV wavelengths that are important for Earth atmospheric studies, and particularly for ozone chemistry.
\end{abstract}



\section{Introduction}

The radiative flux spectrum incident at the top of the Earth atmosphere is known as the solar spectral irradiance (SSI). The near-UV range of the SSI plays a fundamental role in the ozone chemistry. Most of the Earth atmospheric ozone is produced in the stratosphere as a result of photo-dissociation of molecular oxygen by UV Solar radiation in the Schuman-Runge bands at wavelength shorter than $\sim$ 1750 \AA, and destroyed by ozone photo dissociation in the Hartley-Huggins bands (1900 to 3900 \AA) \citep{fon99}. 
A change in the amount of near-UV SSI affects directly the destruction of ozone leading to tropospheric effects. Ozone changes also affect the amount of biologically harmful near-UV radiation reaching the Earth's surface \citep{bra97}. For these reasons, there is a need for an accurate calculation of the solar radiation that reaches the top of the Earth atmosphere in the near-UV range. 

The Solar Radiation Physical Modeling (SRPM) system is a code library that permits one to compute the radiance (emergent intensity from the solar surface) spectrum of the Sun based on physical models of the solar atmosphere (see Fontenla et al. 2006, 2007, and 2009).
The code solves the radiative transfer and non-LTE statistical equilibrium assuming a specified one dimensional plane$-$parallel geometry or a spherical symmetry depending of the study. 
The paper by \citet{fon11} applied the SRPM code and a set of physical models of various features observed on the solar surface to the calculation of high spectral-resolution SSI from extreme ultraviolet to far infrared. 
The results obtained in that paper allow one to study the effects of SSI changes on the Earth and other planetary atmospheres. Comparisons show a generally very good agreement between the observed and synthetic irradiance, but significant differences occur at some wavelengths. 

Important discrepancies, especially in the near-UV range, reflect the need for more accurate and complete atomic models of some important species emitting and absorbing in that range.
Ni I is an important species, considering its abundance in the solar atmosphere, and has strong lines in the near-UV range. The present paper focuses on improving the calculation of the solar spectra in this range by improving the Ni I atomic model and using the SRPM system to compute the non-LTE populations and the emergent intensities. Comparisons are shown between the new results and observations of radiance and irradiance.

Only a few publications concerning Ni I NLTE calculations are found in the literature, most of them about the formation of the Ni I 6769.64  \AA ~line. In helioseismology, this line is used for observing line-of-sight velocity (Doppler shift) by the Michelson Doppler Imager on-board  SOHO and by the ground-based Global Oscillations Network Group. 

\citet{bru93} studied the Ni I 6769.64 \AA ~line using an atomic model with 19 energy levels. The atomic data used for the calculation were that by \citet{Cor81}. The level populations and emergent intensities were computed in NLTE using the VAL C model by \citet{val81}.  This paper confirmed that the line is a good choice for helioseismology.

More recently, \citet{nor06} compared the Ni I 6769.64 \AA ~ and the Fe I 6173 \AA ~lines, both candidates to be used by the Helioseismic and Magnetic Imager (HMI) on-board the Solar Dynamics Observatory (SDO) for observing Doppler velocities and magnetic fields. They adopted a Ni I atomic model with 25 energy levels and the same source of atomic data as in the previously mentioned work. The NLTE calculations were carried out again using model C by \citet{val81} and also using sunspot umbra model M by \citet{mal86}. The HMI team selected the Fe I spectral line because of its better performance for magnetic diagnostics.

Considering the spectral lines and term system of Ni I, the results of \citet{lit93} increased the number of known Ni I levels reported by \citet{Cor81} from 185 to 286, and the number of classified Ni I lines were expanded from 1071 to 1996. This work is the source used by the National Institute of Standards and Technology (NIST) database version 4 \citep{nist11}. The lowest 61 energy levels in the NIST tabulation are adopted in the present paper because they correspond to the most important lines particularly in the near-UV.

In the present work we calculate the Quiet Sun spectrum, as defined by the mix of solar features described by \citet{fon11}, using the same set of physical models but improving the Ni I atomic model from the original 10 to the present 61 levels, and compare the computed spectrum with observations of several Ni I line profiles.

This paper is arranged as follows: the new Ni I atomic model is presented in Section 2. In Section 3 we show some results concerning the radiative transfer and NLTE calculations. In Section 4 we compare the computed quiet sun spectra with observations. Finally, in Section 5 we discuss our results.

\section{The improved Ni I atomic model}

The improvement of the Ni I atomic model is primarily motivated by producing a better computation and comparison with observations of the near-UV SSI. 

Observations in this spectral range are scarce because the atmospheric absorption, especially by ozone, make ground observations impossible. SSI observations from space  usually have limited spectral resolution.

A limitation for computing an accurate spectrum is the availability of accurate atomic data for NLTE calculations.
The NIST atomic spectra database,  \citep{nist11}, lists 503 Ni I spectral lines in the range 1900 to 7000 \AA, with 266 belonging to the near-UV range (1900 to 3900 \AA). 

To reproduce most of the the strong Ni I lines in the near-UV range, it is necessary to consider an atomic model with 61 energy levels, 
and as described by \citet{nist11} the corresponding sublevels defined by the J quantum number. 
Using this number of levels only 34 lines of the NIST listed lines are not included. 

The level energies, line wavelengths, and weighted oscillator strength $gf$ values were taken from the NIST atomic spectra database. The
collisional strength parameters were obtained from the \citet{sea62} approximation, and the photoionization cross sections were estimated 
from similar atomic structure species. The reason for these approximations is the absence of better Ni I atomic data in the literature.
For the radiative, Stark, and Van der Waals broadening parameters, we used the values estimated by \citet{kur}.

In Table 1 we present the description of the 61 atomic levels and sublevels of energy considered. The first 
column  lists the configuration, and the second column the associated term. The third column shows the values of the J quantum number that define the sublevels, and the fourth column presents the sublevel energies. The fifth column lists the level number in our scheme.

\section{The NLTE calculations}

We carried out the calculations using the latest semi-empirical models of the solar atmosphere described by \citet{fon11}. 
The Quiet Sun spectrum was built by weighted averaging of the three models listed in Table 2 and shown in Figure 1. To reproduce 
the low solar activity state during the period 2008-2009, we considered the solar disk as covered by 80\% of internetwork feature (model 
1001, B), 19\% of network feature (model 1002, D) and 1\% of active network (model 1003, F).

The populations for the atomic species listed in Table 2 of \citet{fon11} were all calculated in full NLTE as in \citet{fon11}, but now with the new Ni I atomic model that increased the number of energy levels from 10 to 61. 

We calculated the LTE departure coefficients ($b$ values) defined by the ratio of the computed level populations relative to those 
resulting from the Saha-Boltzman distribution to the next ion stage (Menzel definition, Rutten 2003).
Figure 2 shows the calculated $b$'s for many levels that resulted from the atmospheric model 1001. 

The general behavior is similar to that described in \citet{fon09}. At heights above $\sim$ 900 km (ie, above the temperature minimum region) the $b$ values are greater than unity, with the largest value corresponding to the ground level. 
The opposite occurs in the low chromosphere and the temperature minimum region where the $b$ values are smaller than unity. In general, as the energy of the level increases the departure coefficients are closer to unity, but fluctuations occur depending on how the various levels are connected by permitted transitions.

In Figure 3 we plot the $b_{1}$ departure coefficients of the ground level for the three models used to reproduce the quiet sun spectrum. Although the general behavior is similar, the region where $b_{1}$ is smaller than unity is thinner, and closer to unity, as the minimum temperature of the atmosphere increases.

In Figure 4 we show the ion density fractions of Ni I/Ni$_{total}$, Ni II/Ni$_{total}$, and Ni III/Ni$_{total}$ with height. Throughout the atmosphere the total Ni II density is nearly the total Ni density. Above $\sim$300 km Ni I/Ni$_{total}$ drops slowly. Above $\sim$600 km Ni III/Ni$_{total}$  begins to increase, although below the chromosphere-corona transition region it remains several orders of magnitude lower than the other two density fractions.

In Figure 5, we plot the line source function (solid line) and the local Planck function (doted line) versus the optical depth at line center $\tau$ for four important line transitions. In a first approximation, line center forms at $\tau=1$ for disk center and $\tau=2/3$ for the irradiance. Considering this plot, departures from LTE produce a large difference between these functions for most of the plotted $\tau$ values , and specially around $\tau=1$. 
The heights at which $\tau=1$ are as follows; $\sim 597$ km for the 2747.554 \AA ~line, $\sim 945$ km for the 3415.74 \AA ~line, $\sim 840$ km for the 3567.39 \AA ~line, and $\sim 323$ km for the 6769.64 \AA ~line. Thus, the centers of these lines form in the low chromospheric layers at and below the temperature region where NLTE effects are strong, and their source functions cannot be well described by any simple approximation.

\section{Comparison of the calculated spectra with observations}

We carried out the calculations of the solar spectrum with the new Ni I atomic model, using the latest semi-empirical models of the solar atmosphere described by \citet{fon11}. 
The Quiet Sun spectrum was constructed by weighted averaging of the spectra from three models listed in Table 2 and shown in Figure 1. To reproduce the low solar activity state during the period 2008-2009,  the solar disk is considered covered by the relative areas of features indicated in Table 2.

The information about the Ni I lines and energy levels that we compare with observations are presented in Table 3. The last column in this Table indicates the figure number where this transition is plotted.  Many of the line profiles shown in the Figures correspond to transitions between energy levels that were never before computed in full NLTE. 

Computed spectral lines in the range from 3290 to 7000 \AA ~are compared with the observations at disk-center in the FTS Solar Atlas by \citet{bra99}. The absolute intensities of these observations were from the Kitt Peak FTS-data as described
by Neckel and Labs (1984a, b) and Neckel (1994). The absolute calibration of this disk-center spectrum is not very reliable as explained by \citet{bra99}. For this reason, we scaled the observations to match the continuum near the line profiles with the calculations. 
Figures 6 to 8 show the comparison of several line profiles and the scaling factors.

The last panel in Figure 7 shows the Ni I 6769.64 \AA ~(vaccum wavelength) line used in SOHO/MDI  and GONG instruments. 

The differences between the two computed profiles plotted in these Figures are due to the different number of levels in both NLTE calculations that was mentioned before. Several Ni I lines were absent in the spectra calculated by Fontenla et al. (2011), while others are too deep (e.g. dashed trace in Figure 7). The later are improved in our present calculation by the inclusion of many more levels.

In some cases the calculated line profile is wider than the observed one. We also recomputed the line profiles using the broadening parameters included in the Vienna Atomic Line Database (VALD) \citep{kup00}, but the results obtained were similar. 

Comparisons with observations were also carried out with near-UV irradiance spectrum in the range from 190 to 310 nm. 
The data from space instruments by Thuillier et al. (2003), and by Harder et al. (2010) have a reliable absolute calibration. However, 
these data have low spectral resolution.
In Figure 9 we show the comparison between our computed spectral irradiance and the Composite 3 reference spectrum by \citet{thu03}. 
To simulate the resolution of the observations, the calculated spectral irradiance was convolved with a 1 nm FWHM $cos^{2}$ ~function as in \citet{fon11}.

The observations of the irradiance spectrum by \citet{hal91} and \citet{and89}, hereafter H\&A, were obtained from stratospheric balloon flights that are very difficult to calibrate because of the atmospheric absorption. However, these observations have a fairly good spectral resolution of 0.01 nm.
For the present comparison the H\&A spectrum was scaled to overall match the values in the calculated spectrum. In Figure 10 we compare several computed Ni I line profiles with the scaled H\&A irradiance observations. The comparison shows good agreement between calculations and observations.

\section{Summary}

This work improves the Ni I atomic model to include the majority of the lines in the near UV range that are listed in the NIST 
atomic database. Several of these lines are very strong and have not been calculated in full NLTE by previous papers. 

The results shown here display a fairly good agreement with the observed Ni I lines because we include full NLTE calculations for an atomic model with 61 levels, which have fine structure sublevels. This agreement does not require any changes of the \citet{fon11} atmospheric models.

Also, we reproduce well the Ni I 6769.64 \AA ~used in MDI and GONG observations, allowing a better understanding of the line formation in different solar features, and investigating the impact of solar activity on its behavior. 

The results presented in this work stress the importance of full NLTE calculation and the use of a sufficient number of levels for realistic computation of line profiles. From these results it is clear that using the \citet{fon11} models for spectra calculations with LTE codes will not properly compute the deep Ni I lines and likely others. This is because the calculations here show that the line source function substantially departs from LTE in the low chromosphere, and at the temperature minimum. LTE values will systematically produce wrong results with the \citet{fon11} models.

\acknowledgments

MCV acknowledges support from PICT 0746-2010 ANPCyT, and Programaci\'{o}n Cient\'{i}fica 2010-2011 UNTREF grants. JMF was supported by LWS-NASA grant NNX09AJ22G.

\clearpage

\setcounter{table}{0}
\begin{table}
\begin{center}
\caption{The first 61 levels of Ni I used in the present paper \citep{nist11}. \label{tab1}} 
\begin{tabular}{lcccc} 
\tableline\tableline
 Configuration  & Term  & J  & Level ($cm^{-1}$) & Level number\\ \tableline
 3d$^{8}$($^3$F)4s$^{2}$  & $^3$F  & 4  &  0.000 & 1 \\ 
      & & 3   & 1 332.164 &   \\
      & & 2  &  2 216.550 & \\   
3d$^9$($^2$D)4s  & $^3$D  & 3  &  204.787 & 2 \\  
   & & 2  & 879.816  &  \\
   & & 1  &  1 713.087 &  \\
3d$^9$($^2$D)4s & $^1$D  & 2  &  3 409.937 & 3  \\
3d$^8$($^1$D)4s$^2$  &  $^1$D  & 2  & 13 521.347 & 4  \\
3d10  & $^1$S  & 0  &  14 728.840 & 5  \\
3d$^8$($^3$P)4s$^2$  & $^3$P  & 2  &  15 609.844 &  6 \\
         &  & 1    &  15 734.001 & \\
          & & 0  &  16 017.306 & \\ 
3d$^8$($^1$G)4s$^2$  & $^1$G  & 4  & 22 102.325 & 7\\
3d$^8$($^3$F)4s4p($^3$P$^o$)  & $^3$D$^o$  & 4  & 25 753.553 & 8\\
               &   & 3  &  26 665.887 & \\
   & & 2  &  27 414.868 & \\
&  & 1  &  27 943.524 & \\
 & & 0  &  28 212.998 & \\
3d$^8$($^3$F)4s4p($^3$P$^o$)  & $^5$G$^o$  & 6  &  27 260.894 & 9\\
 & & 5  &  27 580.391 &\\
 & & 4  &  28 068.065 & \\
 & & 3  &  28 578.018 & \\
 & & 2  &  29 013.206 & \\
\tableline 
 \end{tabular}
 \end{center}
\end{table}

\clearpage

\setcounter{table}{0}
\begin{table}
\begin{center}
\caption{The first 61 levels of Ni I used in the present paper \citep{nist11}. \label{taba}} 
\begin{tabular}{lcccc} 
\tableline\tableline
 Configuration  & Term  & J  & Level ($cm^{-1}$) & Level number\\ \tableline
3d$^8$($^3$F)4s4p($^3$P$^o$)  & $^5$F$^o$  & 5  & 28 542.105 & 10 \\
  & &4  &  29 084.456 &\\
  & & 3  &29 832.779 &  \\
  & & 2  &  30 163.124 & \\
  & & 1  &  30 392.003 & \\
3d$^9$($^2$D)4p  & $^3$P$^o$  & 2  &  28 569.203 & 11\\
  & & 1  & 29 500.674 & \\
  & & 0  & 30 192.251 & \\
3d$^9$($^2$D)4p  & $^3$F $^o$ &4  & 29 480.989 & 12 \\
  & & 3  & 29 320.762 &  \\  
  & & 2  &30 619.414 & \\
3d$^9$($^2$D)4p  & $^3$D$^o$ & 3 &  29 668.893 & 13\\
 &  & 2  & 29 888.477 & \\
&  & 1  &  30 912.817 & \\
3d$^9$($^2$D)4p  & °  & 3  &  29 668.918 & 14 \\
3d$^8$($^3$F)4s4p($^3$P$^o$)  & $^3$G$^o$  & 5  & 30 922.734 &  15 \\
 &  & 4  & 30 979.749 & \\
&  & 3  & 31 786.162 & \\ 
3d$^9$($^2$D)4p  & $^1$F$^o$ & 3  &  31 031.020 & 16 \\
\tableline 
 \end{tabular}
 \end{center}
\end{table}

\clearpage

\setcounter{table}{0}
\begin{table}
\begin{center}
\caption{Continuation \label{taba}} 
\begin{tabular}{lcccc} 
\tableline\tableline
 Configuration  & Term  & J  & Level ($cm^{-1}$) & Level number\\ \tableline
3d$^9$($^2$D)4p  & $^1$D$^o$ & 2  & 31 441.635 & 17\\
3d$^8$($^3$F)4s4p($^3$P$^o$) & $^3$F$^o$  & 4  &  32 973.376 & 18\\
&   & 3  & 33 112.334 & \\
 &  & 2  &  34 163.264 & \\
3d$^9$($^2$D)4p  & $^1$P$^o$  & 1  & 32 982.260 & 19\\
3d$^8$($^3$F)4s4p($^3$P$^o$)  & $^3$D$^o$  & 3  &  33 500.822 & 20 \\
 &  & 2   & 33 610.890 & \\ 
 &  & 1  &  34 408.555 & \\
3d$^8$($^3$F)4s4p($^3$P$^o$) & $^1$G$^o$ & 4 &  33 590.130 & 21\\  
3d$^8$($^3$F)4s4p($^3$P$^o$)  & $^1$F$^o$ & 3 &  35 639.122 & 22\\  
3d$^8$($^3$F)4s4p($^3$P$^o$) & $^1$D$^o$ & 2  &  36 600.791& 23\\  
3d$^8$($^3$P)4s4p($^3$P)  & $^5$P$^o$ & 3 &  40 361.249 &  24\\  
 & & 2  &  40 484.212 & \\ 
  & & 1  & 40 768.996 & \\
3d$^8$($^1$D)4s4p($^3$P$^o$) & $^3$F$^o$ & 4  & 42 585.212 & 25\\
 & & 3  &  42 767.853 & \\  
 & & 2  &  42 954.203 & \\
3d$^9$($^2$D$_{5/2}$)5s & $^2$[5/2]  & 3  &  42 605.945 & 26\\
  & & 2  & 42 790.010 & \\   
3d$^8$($^1$D)4s4p($^3$P$^o$) & $^3$D$^o$ & 3  &  42 620.994 & 27 \\
  & & 2  & 42 653.661 & \\
 & & 1  &  42 656.289 & \\
3d$^8$($^3$F)4s4p($^1$P$^o$)  & $^3$G$^o$ & 5  &  43 089.578 & 28 \\
 & & 4 &  44 314.904 &  \\  
 & & 3  &  44 565.037 & \\  
\tableline 
 \end{tabular}
 \end{center}
\end{table}

\clearpage

\setcounter{table}{0}
\begin{table}
\begin{center}
\caption{Continuation \label{tabb}} 
\begin{tabular}{lcccc} 
\tableline\tableline
 Configuration  & Term  & J  & Level ($cm^{-1}$) & Level number\\ \tableline
3d$^8$($^3$F)4s4p($^1$P$^o$)  & $^3$F$^o$ & 4  &  43 258.726 & 29\\
 & & 3 &  45 281.089 &    \\
 & & 2  & 45 418.804 &  \\
3d$^8$($^1$D)4s4p($^3$P$^o$) & $^3$P$^o$ & 1  &  43 463.981  & 30\\
 & & 2  & 43 933.408&    \\
3d$^8$($^3$F)4s4p($^1$P$^o$) & $^3$D$^o$ & 3  &  43 654.903 & 31 \\
  & & 2  & 44 475.099 &  \\ 
 & & 1  &  45 122.383 &  \\
3d$^8$($^3$F)4s4p($^1$P$^o$)  & °  & 3 &  43 654.974 & 32 \\
3d$^8$($^3$P)4s4p($^3$P$^o$) & $^5$D$^o$ & 4 &  44 540.525  & 33 \\
 & & 3  &  44 206.099 &  \\
 & & 2 &  44 093.773  &  \\
 & & 1  &  44 132.250  & \\
 & & 0  &  44 414.955 &  \\
3d$^9$($^2$D$_{3/2}$)5s  & $^2$[3/2] & 1 & 44 112.173 & 34  \\
 & & 2 &  44 262.599 &  \\
3d$^8$($^3$P)4s4p($^3$P$^o$)  & ° & 4 &  44 336.10 & 35  \\
3d$^8$($^3$P)4s4p($^3$P$^o$) & $^3$P$^o$& 2  &  46 522.866 & 36 \\
 & & 1 &  47 208.149 & \\  
 & & 0  &  47 686.587  &\\ 
\tableline 
 \end{tabular}
 \end{center}
\end{table}

\clearpage

\setcounter{table}{0}
\begin{table}
\begin{center}
\caption{Continuation \label{tabc}} 
\begin{tabular}{lcccc} 
\tableline\tableline
 Configuration  & Term  & J  & Level ($cm^{-1}$) & Level number\\ \tableline
3d$^8$($^3$P)4s4p($^3$P$^o$) & $^3$D$^o$& 3 & 47 030.102 & 37 \\
 & & 2  &  47 139.337 &  \\   
  & & 1 & 47 424.785  & \\
3d$^8$($^3$P)4s4p($^3$P$^o$) & $^5$S° & 2 &  47 328.784 & 38  \\
3d$^8$4s($^4$F)5s  & $^5$F & 5  & 48 466.490 & 39  \\
& & 4  &  49 085.982 & \\
&  & 3 &  49 777.569 &  \\
&  & 2 & 50 346.427 &  \\
&  & 1 &  50 744.552 &\\
3d$^9$($^2$D)5p  & $^1$F$^o$ & 3  &48 671.049 & 40  \\
3d$^9$($^2$D)5p & $^3$F$^o$& 4  &  48 715.586  & 41 \\
 & & 3  &  50 142.991 &   \\
 & & 2 &  50 039.191 &  \\
3d$^9$($^2$D)5p & $^3$P$^o$& 2 &  48 735.290 & 42  \\
  & & 1 &  49 196.181 &  \\
  & & 0 &  50 138.458 &  \\
3d$^8$($^3$P)4s4p($^3$P$^o$)  & $^1$P$^o$ & 1 &  48 818.097  & 43 \\
3d$^9$($^2$D$_{5/2}$)4d  & $^2$[1/2] & 1 &  48 953.316  & 44\\ 
 & & 0  & 49 610.345 &  \\  
3d$^9$($^2$D)5p & $^1$D$^o$& 2  &  49 032.926 & 45  \\
3d$^9$($^2$D$_{5/2}$)4d & $^2$[9/2] & 5 &  49 158.480  & 46 \\
  & & 4 & 49 174.770 & \\  
3d$^9$($^2$D$_{5/2}$)4d & $^2$[3/2] & 2  &  49 159.030  & 47\\
  & & 1  & 49 171.151 &   \\

\tableline 
 \end{tabular}
 \end{center}
\end{table}

\clearpage

\setcounter{table}{0}
\begin{table}
\begin{center}
\caption{Continuation \label{tabd}} 
\begin{tabular}{lcccc} 
\tableline\tableline
 Configuration  & Term  & J  & Level ($cm^{-1}$) & Level number\\ \tableline
3d$^8$($^3$P)4s4p($^3$P$^o$) & $^1$D$^o$& 2  &  49 185.138 & 48  \\
3d$^9$($^2$D$_{5/2}$)4d & $^2$[5/2] & 3 &  49 271.540  & 49 \\
 &  & 2 & 49 327.811  & \\
3d$^9$($^2$D$_{5/2}$)4d & $^2$[7/2] & 3  &  49 313.814  & 50 \\
  &  & 4 & 49 332.593  & \\
3d$^9$($^2$D)5p  & $^3$D$^o$& 3 &  49 328.140 & 51 \\
 & & 2 & 50 689.489 &   \\
 & & 1 &  50 851.199 &  \\
3d$^8$($^3$P)4s4p($^3$P$^o$) & $^3$S$^o$ & 1 & 49 403.386 & 52  \\
3d$^9$($^2$D)5p  & ° & 3 &  50 142.8  & 53   \\
3d$^8$($^1$S)4s$^2$ & $^1$S & 0 &  50 276.321 & 54  \\
3d$^9$($^2$D)5p  & $^1$P$^o$& 1 &  50 458.192  & 55 \\
3d$^8$4s($^4$F)5s & $^3$F  & 4 & 50 466.131  & 56 \\
  & & 3  & 51 306.038 &  \\
  & & 2  &  52 040.523 & \\ 
3d$^9$($^2$D$_{3/2}$)4d  & $^2$[1/2] & 1 &  50 536.703 & 57 \\
  & & 0 & 51 457.250  &  \\
3d$^9$($^2$D$_{3/2}$)4d & $^2$[7/2] & 3  &  50 677.555 & 58 \\
  & & 4  & 50 706.273 &  \\  
3d$^9$($^2$D$_{3/2}$)4d  & $^2$[3/2]  & 1 &  50 716.896  &  59\\
 & & 2 &  50 754.103 & \\  
3d$^8$($^1$G)4s4p($^3$P$^o$) & $^3$F$^o$& 4 &  50 789.303  & 60 \\
  & & 3  & 51 124.662  & \\
 & & 2  &  51 343.547 & \\ 
3d$^9$($^2$D$_{3/2}$)4d & $^2$[5/2]  & 3  &  50 832.001 & 61\\
 & & 2  &  50 834.401 & \\
\tableline 
 \end{tabular}
 \end{center}
\end{table}

\clearpage

\begin{table}
\begin{center}
\caption{Quiet Sun feature components and their respective models. We follow the feature and model index designations as in Fontenla et al. (2011). \label{tab2}} 
\begin{tabular}{lccc} 
\tableline\tableline
 Feature &  Description & Photosphere-Chromosphere &  Relative area on \\
     &  &  Model Index  &  the Solar Disk\\ \tableline
 B & Quiet Sun inter-network & 1001 & 80~\% \\ 
 D & Quiet Sun network lane & 1002 & 19~\% \\ 
 F & Enhanced network & 1003 & 1~\% \\ 
\tableline 
 \end{tabular}
 \end{center}
\end{table}

\clearpage
\begin{table}
\begin{center}
\caption{Transition lines plotted in Figures 6, 7, 8 and 10. In column 1, the vacuum wavelength. In column 2 and 3, the energy value for the upper and lower sublevel respectively involved in the transition \citep{nist11}. \label{tab3}} 
\begin{tabular}{lccc} 
\tableline\tableline
 $\lambda_{vac}(\AA)$ &  Upper level $(cm^{-1})$& Lower level $(cm^{-1})$ & Figure number \\ \tableline
 2296.69 & 43654.974 & 0 & 10 \\ 
 2326.515 & 44314.904 & 1332.164 & 10 \\ 
 2747.554 & 36600.791 & 204.787 & 10 \\ 
2822.122 & 35639.122& 204.787& 10\\
2982.516 & 34408.555& 879.816 & 10\\
3003.36  & 33500.822 & 204.787 & 10\\
3370.531 & 29668.918 & 0 & 6\\
3393.957& 29668.918 & 204.787&6\\
3415.744 & 29480.989& 204.787 &6\\
3434.54 & 29320.762& 204.787 &6\\
3447.247 & 29888.477 & 879.816 & 6\\
3459.451 & 30619.414& 1713.087 & 6\\
3493.956 & 29500.674 & 879.816 & 7\\
3516.057 & 29320.762 & 879.816 & 7\\
3525.544 & 28569.203 & 204.787 & 7\\
3567.39 & 31441.635 & 3409.937& 7\\
3859.391 & 29320.762 & 3409.937 & 7\\
4717.081 & 49777.569& 28578.018 & 7\\
4830.372 & 49271.540 & 28569.203 & 8\\
4919.737 & 51306.038 & 30979.749 & 8\\
5036.766 & 49174.770 & 29320.762 & 8\\
5081.949 & 49158.480 & 29480.989 & 8\\
5478.426 & 32982.260 & 14728.84 & 8\\
6769.64 & 29500.674 & 14728.84 & 8\\
\tableline 
 \end{tabular}
 \end{center}
\end{table}

\clearpage

\begin{figure}
\begin{center}
 \includegraphics[width=6.in]{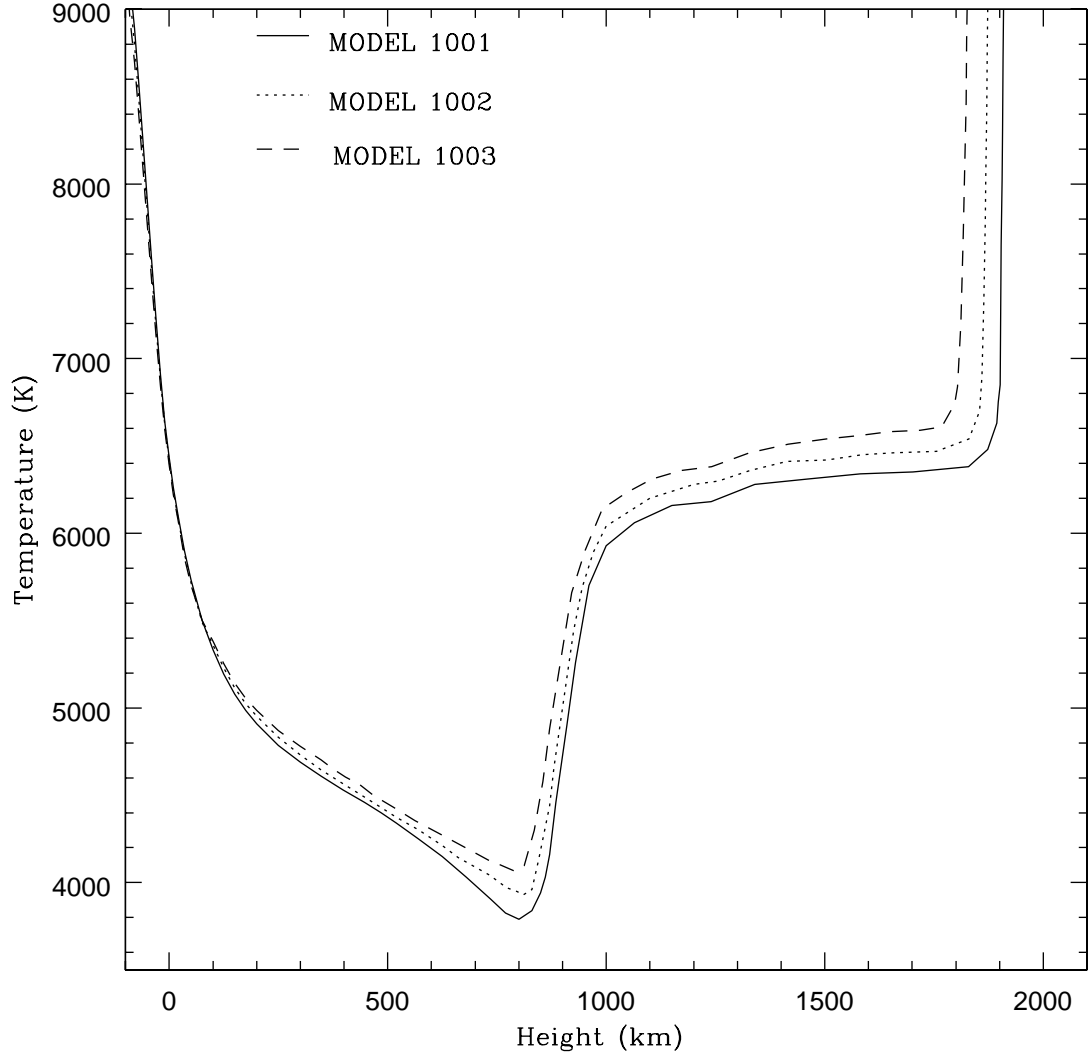} 
 \caption{Atmospheric models for the solar features used to calculate the quiet sun spectrum. These models were built by Fontenla et al. (2011)}
   \label{fig1}
\end{center}
\end{figure}

\clearpage

\begin{figure}
\begin{center}
 \includegraphics[width=5.in]{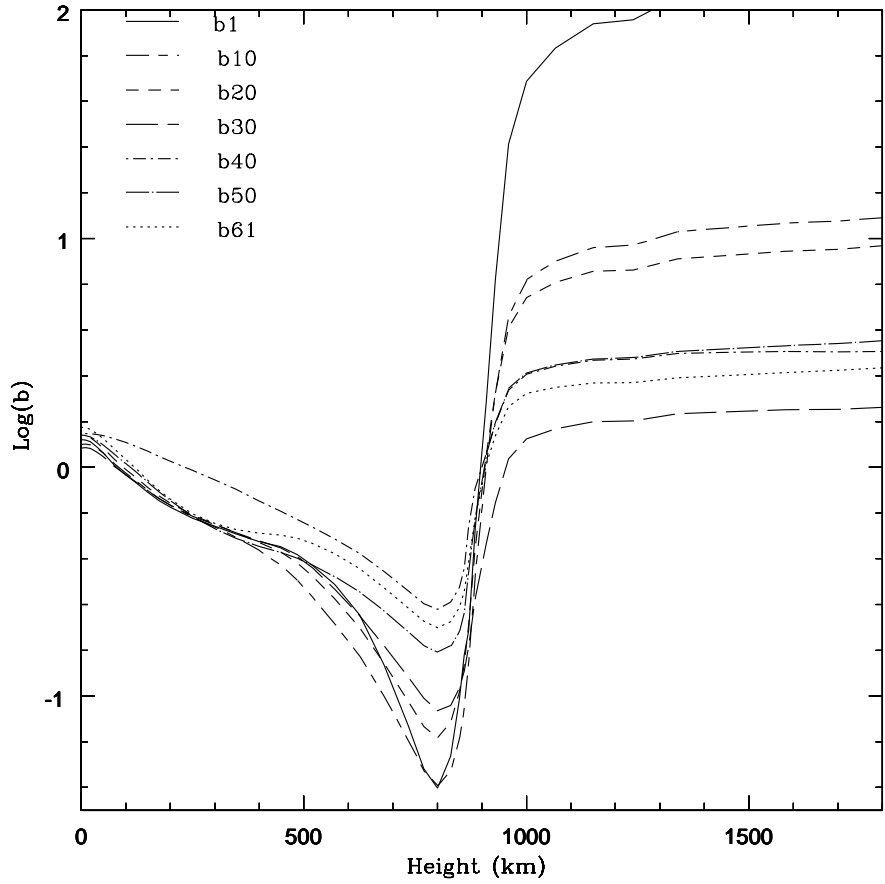} 
 \caption{Departure coefficients $b$ for selected levels of Ni I in the atmospheric model 1001.}
   \label{fig2}
\end{center}
\end{figure}

\clearpage

\begin{figure}
\begin{center}
 \includegraphics[width=5.in]{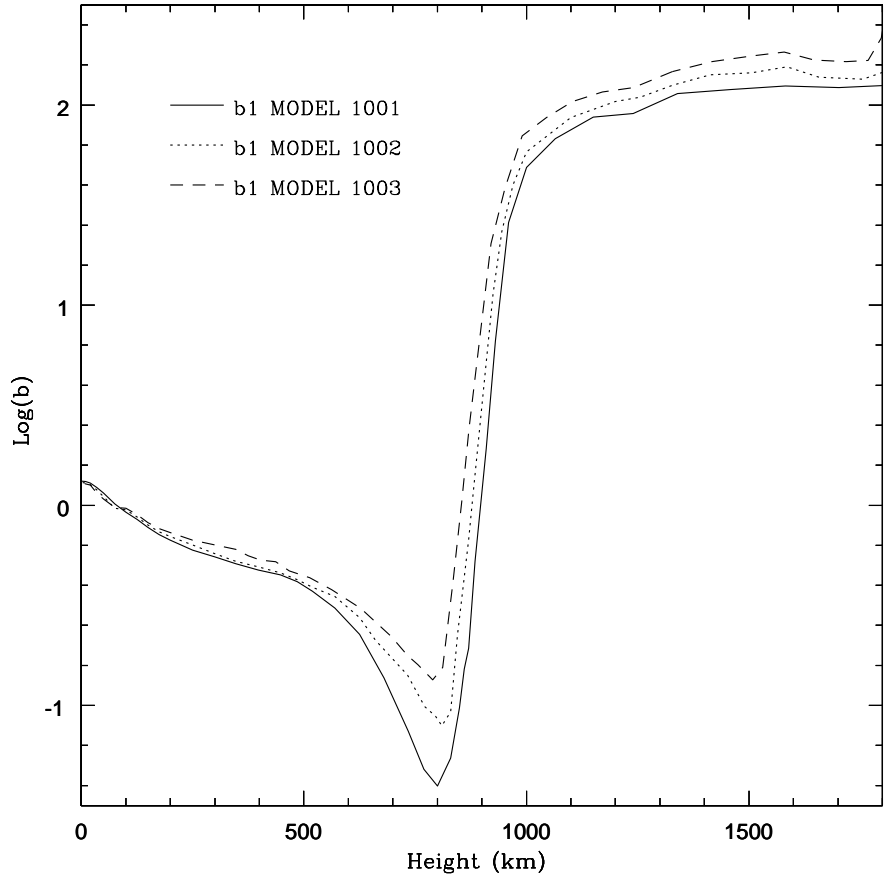} 
 \caption{Departure coefficients $b1$ for the ground state of Ni I in the atmospheric models used to calculate the solar spectrum}
   \label{fig3}
\end{center}
\end{figure}

\clearpage

\begin{figure}
\begin{center}
 \includegraphics[width=5.in]{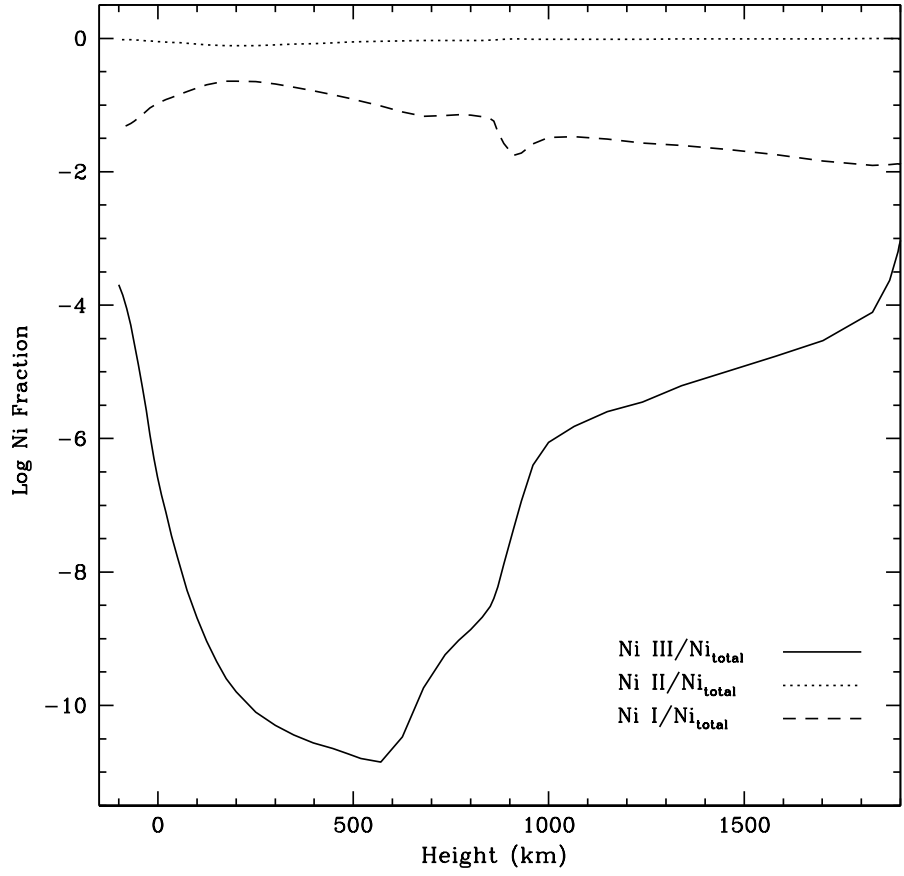} 
 \caption{ Ni I, Ni II and Ni III densities relative to Ni$_{total}$ density for the atmospheric model 1001 as a function of height.}
   \label{fig4}
\end{center}
\end{figure}

\clearpage

\begin{figure}
\begin{center}
 \includegraphics[width=6.5in]{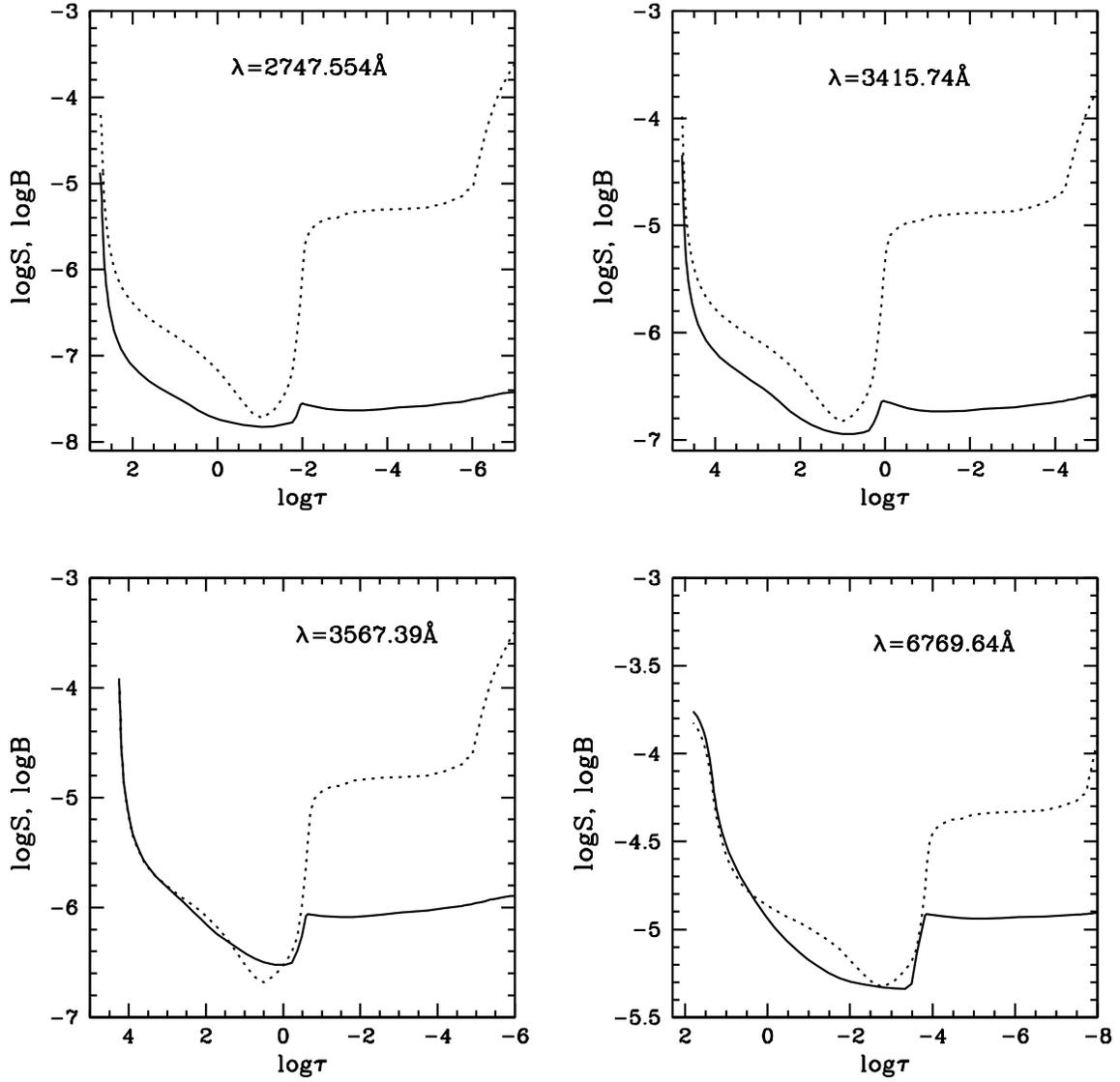} 
 \caption{Line Source Function ($S$, solid line) and Planck Function ($B$, doted line) as a function of the optical depth at line center,$\tau$, for selected spectral lines.}
   \label{figb}
\end{center}
\end{figure}

\clearpage

\begin{figure}
\begin{center}
 \includegraphics[width=6.5in]{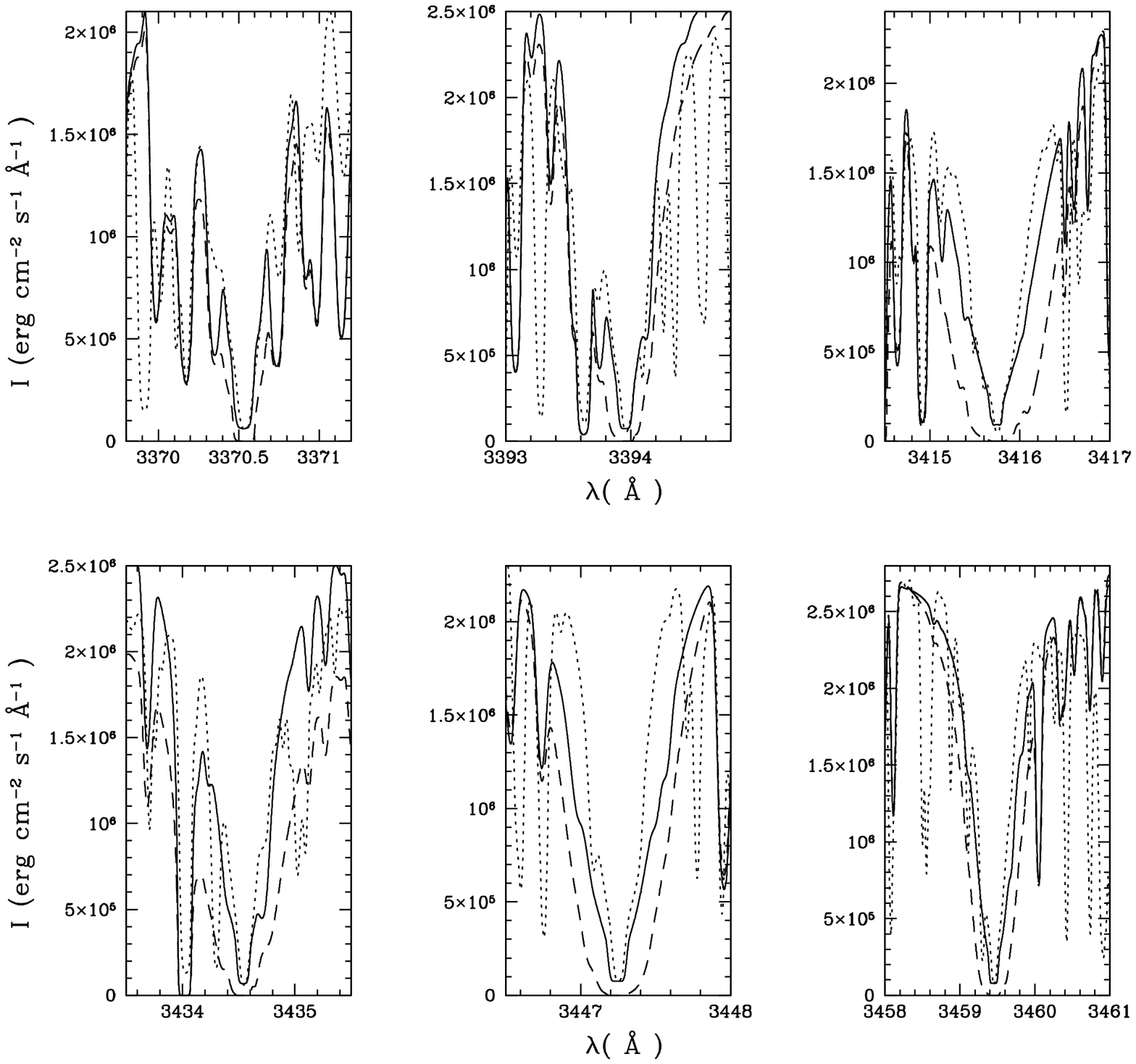} 
 \caption{Comparison between observed and synthetic line profiles at disk center. Solid lines show our calculation, dashed lines the spectra calculated by Fontenla et al. (2011), and doted lines the observations by \citet{bra99}. The scale factors, in increasing wavelength order, are 1, 1, 0.91, 1, 1.18, 1.18. All wavelengths are in vacuum. }
   \label{fig5}
\end{center}
\end{figure}

\clearpage

\begin{figure}
\begin{center}
 \includegraphics[width=6.5in]{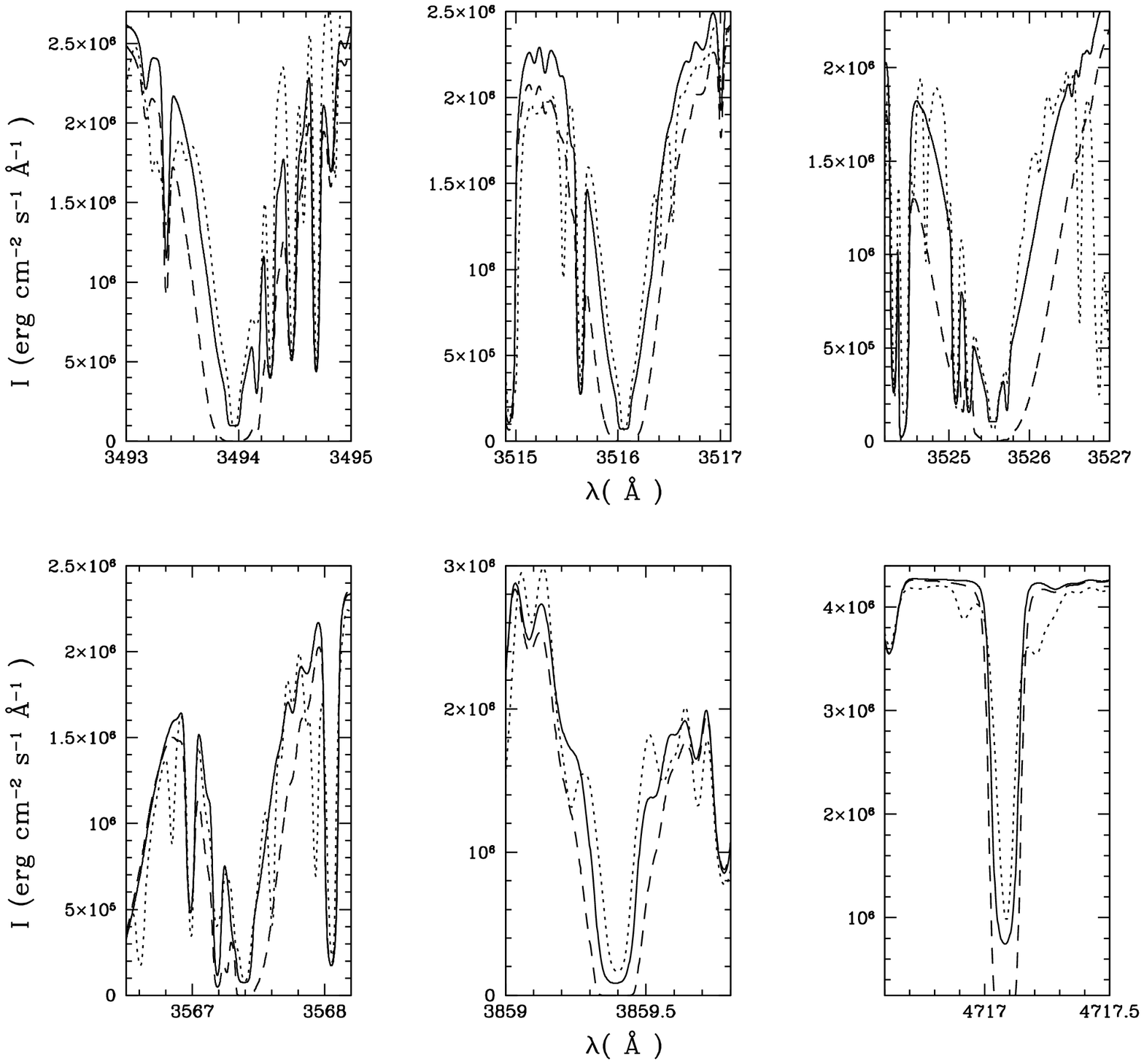} 
 \caption{Comparison between observed and synthetic line profiles at disk center. Solid lines show our calculation, dashed lines the spectra calculated by Fontenla et al. (2011), and doted lines the \citet{bra99} Solar Atlas. The scale factors, in increasing wavelength order, are 1.28, 1, 0.91, 1, 1.18, 1.25. All wavelengths are in vacuum.}
   \label{fig6}
\end{center}
\end{figure}

\clearpage

\begin{figure}
\begin{center}
 \includegraphics[width=6.5in]{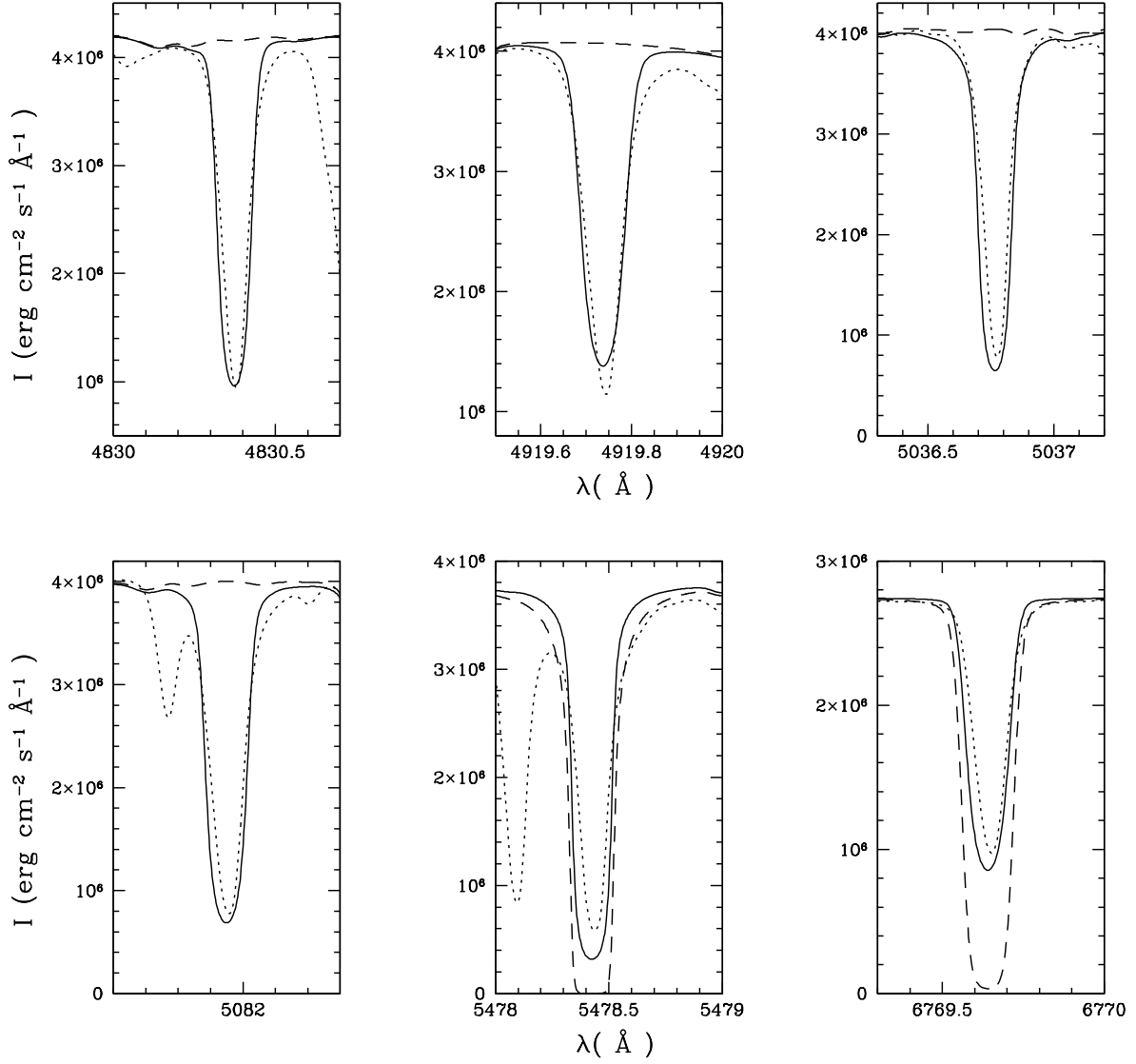} 
 \caption{Comparison between observed and synthetic line profiles at disk center. Solid lines show our calculation, dashed lines the spectra calculated by Fontenla et al. (2011), and doted lines the \citet{bra99} Solar Atlas. The scale factor for all the lines is 1.28. All wavelengths are in vacuum.}
   \label{fig7}
\end{center}
\end{figure}

\clearpage

\begin{figure}
\begin{center}
 \includegraphics[width=6.5in]{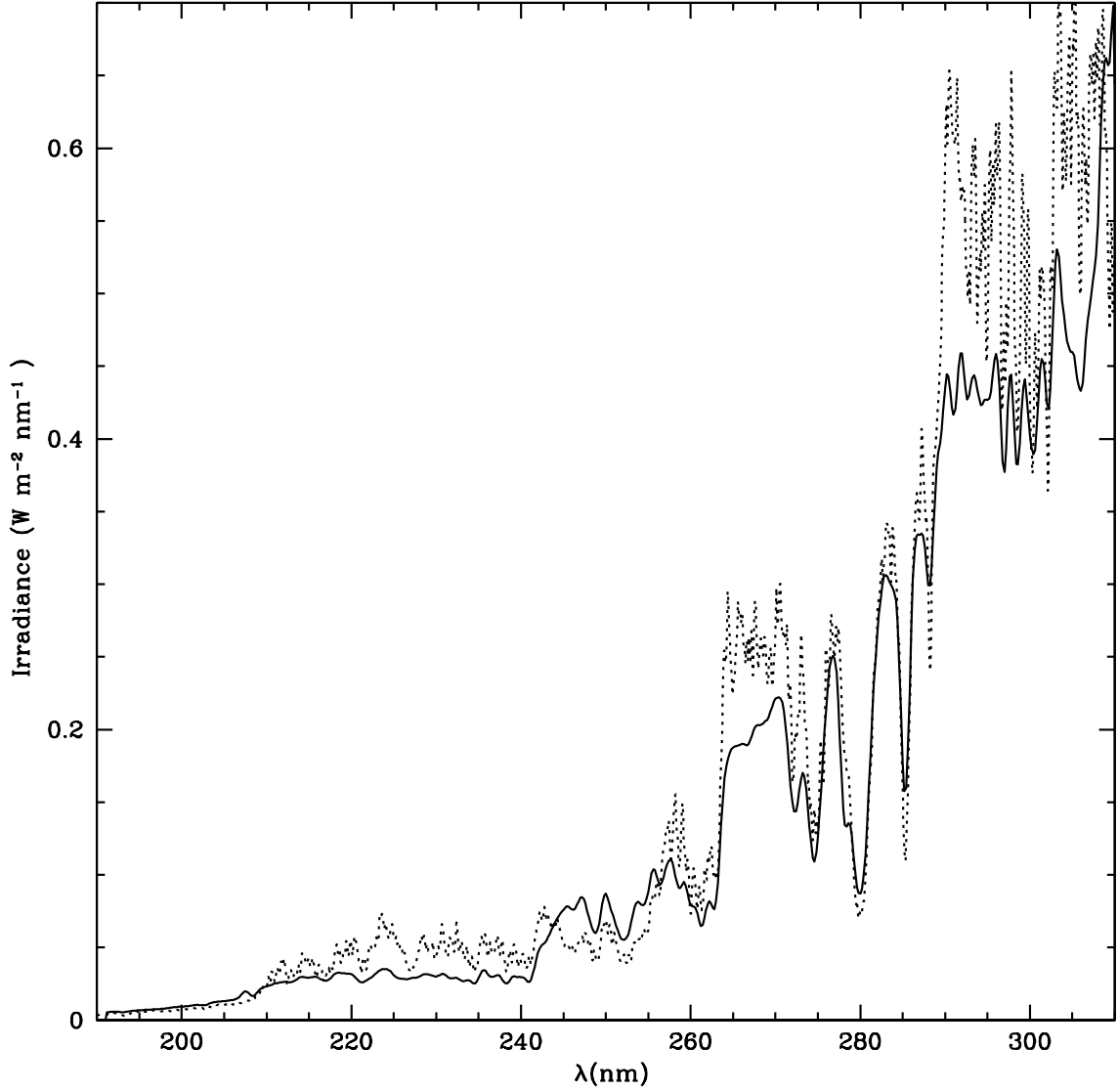} 
 \caption{Comparison between observed and synthetic spectral irradiance between 190 and 310 nm. Solid lines show our calculation, dashed lines the Composite 3 reference spectrum derived by \citet{thu03}}
   \label{fig8}
\end{center}
\end{figure}

\clearpage

\begin{figure}
\begin{center}
 \includegraphics[width=6.5in]{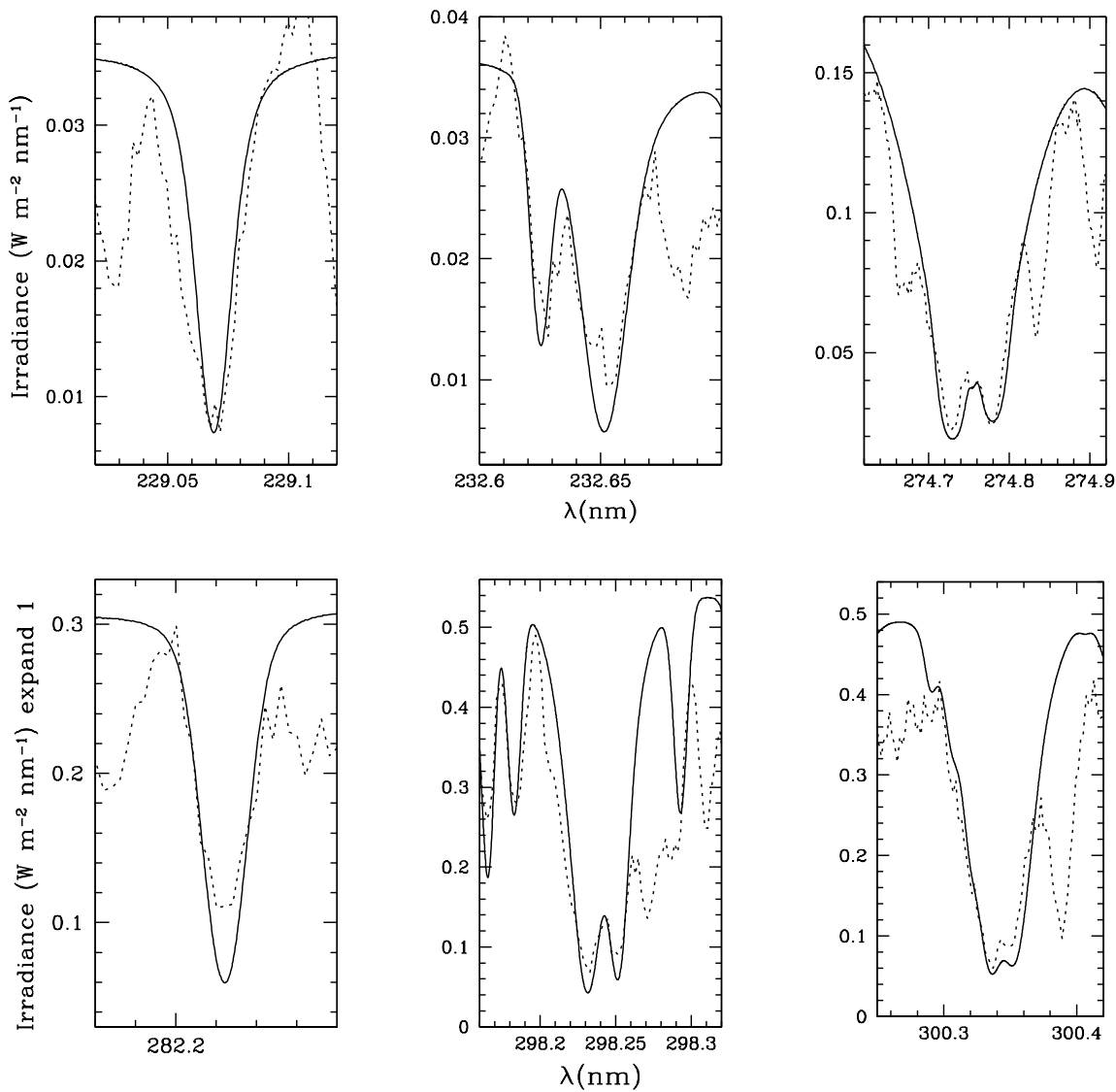} 
 \caption{Comparison between observed and synthetic irradiance line profiles. Solid lines show our calculation, and doted lines the H\&A observations. All wavelengths are in vacuum. The scale factors  between the observed and the calculated profile are, in increasing wavelength order, 1.9, 3.3, 3.5, 1.8, 0.75 and 0.83}
   \label{fig9}
\end{center}
\end{figure}

\clearpage

\end{document}